\documentstyle[aps,prl,epsfig]{revtex}

\begin{document}

\title{Kronecker delta energy terms in thermal field theory}

\author{F. T. Brandt$^a$, Ashok Das$^b$, 
J. Frenkel$^a$ and  J. C. Taylor$^c$}
\address{$^a$Instituto de F\'{\i}sica,
Universidade de S\~ao Paulo,
S\~ao Paulo, SP 05315-970, BRAZIL}
\address{$^b$Department of Physics and Astronomy,
University of Rochester,
Rochester, NY 14627-0171, USA}
\address{$^c$ Department of Applied Mathematics 
and Theoretical Physics,
University of Cambridge,
Cambridge, UK}

\maketitle
\vskip .5cm

\begin{abstract}

We discuss, using the imaginary time method, some aspects of the
connection between the Ward identity, the non-analyticity of
amplitudes and the causality relation in QED at finite temperature.  
\end{abstract}

\section{Introduction}
 
It is well known that, at finite temperature, field theoretic
amplitudes (Greens functions) depend on both external energy and
momentum independently. Furthermore, it is also known that new branch
cuts develop in a thermal field theory due to the existence of new
channels. As a consequence of these new cuts, amplitudes become
non-analytic at the origin in the energy-momentum plane and the limits
$k_{0}\rightarrow 0$ and $|\vec{k}|\rightarrow 0$ do not commute in
general \cite{weldon,kobes,evans,weldon1}. In coordinate space, this simply
means  that the thermal effects are basically nonlocal
\cite{braaten,frenkel}. 

In an earlier paper \cite{das}, the behavior
of thermal
perturbative amplitudes in a $0+1$ dimensional gauge theory was
studied in detail and the
results (which also generalize to other $0+1$ dimensional gauge
models \cite{barcelos}) are
quite interesting. For example, it was found that the self-energy
(this is also true for $n$-point amplitudes) vanishes when the external
energy is non-vanishing, while it is nonzero for a vanishing external
energy. Explicitly, one obtains
\begin{equation}
\Pi^{(0+1)} (k_{0})  = e^{2}\,{d\over dm}\,\tanh {m\over
2T}\,\delta_{k_{0},0} 
\end{equation}
Such a  behavior is, in fact, required by the Ward
identity \cite{das}. Nevertheless, it is quite interesting and
surprising in that 
one would not expect a non-analytic behavior in a $0+1$ dimensional
theory.

In this short note, we analyze, in the imaginary time
formalism,  the origin of such Kronecker
delta terms and show that such a behavior naturally arises in all
dimensions and is a consequence of the analytic continuation of the
external energy to the Minkowski space. We show that the particular
analytic  continuation, which is
conventionally used to ensure good high energy behavior
\cite{baym}, is also the
one  required by the gauge invariance of
the theory and this, in turn, leads to the non-analyticity in the
amplitudes. We show in $3+1$ dimensional QED that these
Kronecker delta terms do have a direct physical meaning. The
organization of the paper is as follows. In section {\bf II}, we analyze
the origin of the non-analyticity in the $0+1$ dimensional QED. This
is a simple model where many aspects of this problem can be clearly
seen. In particular, the analytic continuation that is consistent with
the Ward identity is quite clear in this simple model. In section {\bf
III}, we generalize the results of the $0+1$ dimensional model to QED
in $3+1$ dimensions in the long wave limit. Here, there is a complete
parallel of the Kronecker delta terms, which can be shown to be related to
the electric and the plasmon masses. In section {\bf IV}, we discuss
QED in  $3+1$
dimensions away from the long wave limit and show that the dispersion
relations, to be consistent with the Ward identity, must have a
subtraction at finite temperature. We derive a simple subtracted
dispersion relation for the Kronecker delta terms, which shows that
these are associated with the space-like branch cuts that exist only
at finite temperature. Some of these features are known in the
literature from
earlier studies on the non-analyticity of thermal amplitudes, and  we
have tried, in this note, to present what is known as well as some new
results from a unified point of view. 

\section{$0+1$ dimensional model}

The simplest example where a Kronecker delta term makes its presence
manifest is in the $0+1$ dimensional QED. This model also captures all the
essential characteristics of higher dimensional models. Therefore, let
us discuss this in some detail.

Let us consider a massive fermion, in $0+1$ dimension, interacting
with an Abelian gauge field described by the Lagrangian
\begin{equation}
L =  \overline{\psi} (i\partial_{t} - m - eA)\psi
\end{equation}
where $A$ is the $0+1$ dimensional photon field and let us analyze the
radiative correction to the photon self-energy due to the fermions at
finite temperature. We will use the imaginary time formalism
\cite{kapusta,lebellac,das1} 
throughout  our discussions. We recall that the Ward identities take a very
simple form in $0+1$ dimensions. In particular, for the self-energy,
we have
\begin{equation}
k_{0} \Pi(k_{0}) = 0\label{ward}
\end{equation}
which implies that
\begin{equation}
\Pi(k_{0}) = \delta_{k_{0},0} \Delta
\end{equation}
This, in fact, is the simplest manifestation of Kronecker delta terms
and is a direct consequence of the non-analyticity of amplitudes at finite
temperature. 

It may seem odd that there would be a non-analyticity in a $0+1$
dimensional model, where the only kinematical variable is energy. So,
to understand this better, let us analyze how the Kronecker delta term
arises in a calculation. Let us note that, in the imaginary time
formalism, the photon self-energy is given by
\begin{eqnarray}
\Pi(k_{0}) & = & -e^{2} T \sum_{n} {i((2n+1)\pi T + k_{0})+m\over
((2n+1)\pi T + k_{0})^{2} + m^{2}}\, {i (2n+1)\pi T + m\over
(2n+1)^{2}\pi^{2} T^{2} + m^{2}}\nonumber\\
 & = & - e^{2} T \sum_{n} {1\over -i
((2n+1)\pi T + k_{0}) + m}\,{1\over -i (2n+1)\pi T + m}\label{0+1}
\end{eqnarray}
The sum can be evaluated in a standard manner by noting that
\begin{equation}
T \sum_{n} f(p=i(2n+1)\pi T) = - \sum_{\rm residues} f(p) \tanh
{\beta p\over 2}\label{sum} 
\end{equation}
where $\beta = {1\over T}$ in units of the Botzmann constant and  the
sum, on the right hand side, is  over the residues of the poles of
$f(p)$.  However,
before evaluating the sum, let us note that the behavior of the result is
different depending on whether $k_{0}\neq 0$ or $k_{0} = 0$. When
$k_{0}\neq 0$, we note that $f(p)$ will have two distinct poles,
while for $k_{0} = 0$, the two poles will coincide and we will have a
double pole. Basically, this is the reason for the non-analyticity in
the amplitude and
let us see this in some detail.

Let us note that, for $k_{0}\neq 0$, the evaluation of the sum in
(\ref{0+1})  gives
\begin{equation}
\Pi(k_{0}) = {e^{2}\over 2i k_{0}} \left(\tanh {\beta m\over 2} - \tanh {\beta
(m-ik_{0})\over 2}\right) + (k_{0} \leftrightarrow - k_{0})\label{analytic}
\end{equation}
It is worth noting here that if we assume, at this point, 
that the amplitude has already been analytically continued in $k_{0}$,
then, this will give a non-vanishing value for $\Pi(k_{0})$ for
$k_{0}\neq 0$, which would be inconsistent with the Ward
identity in (\ref{ward}). On the other hand, if we assume that the
external energy  is not 
analytically continued yet and use the fact that $k_{0} = 2\ell\pi T$,
then, it follows from the periodicity of $\tanh {\beta m\over 2}$ that 
\begin{equation}
\Pi(k_{0}) = 0,\qquad k_{0}\neq 0
\end{equation}
which is completely consistent with the Ward identity. This also gives
an additional reason for the analytic continuation, which is
conventionally carried out, in evaluating imaginary time amplitudes.

On the other hand, if $k_{0} = 0$, as we have noted, there is a
double pole and a direct evaluation of the sum, in (\ref{0+1}), using
the  formula in (\ref{sum}) gives
\begin{equation}
\Pi(k_{0} = 0) = e^{2}\, {d\over dm} \tanh {\beta m\over 2} =
\Delta\label{static}
\end{equation}
Mathematically, we can write
\begin{equation}
\Pi (k_{0}) = \widetilde{\Pi} (k_{0}) + \delta_{k_{0},0} \Delta
\end{equation}
where $\widetilde{\Pi}(k_{0})$ is the analytic part of the amplitude
$\Pi(k_{0})$, and this brings out the non-analytic nature of this
expression. Alternatively, we can define
\begin{equation}
\Pi(k_{0} =0) - \lim_{k_{0}\rightarrow 0} \Pi(k_{0}) = \Delta
\end{equation}
which would represent the non-analyticity in the self-energy. Note
that, this non-analyticity can be put into the usual framework by
noting that the $0+1$ dimensional theory can be thought of as the
$\vec{k}=0$ limit of a higher dimensional theory so that the Kronecker
delta term simply corresponds, from the point of view of a higher
dimensional theory (Note that, in a higher dimensional theory, there
will be integration over internal spatial momentum, which does not
occur in the $0+1$ dimensional theory), to
\begin{equation}
\Delta = \Pi(k_{0} = 0, \vec{k}=0) - \lim_{k_{0}\rightarrow 0}
\Pi(k_{0},\vec{k}=0)
\end{equation}
which is precisely the difference between the static limit and the
long wave limit of the self-energy. We will show later that, in QED,
this has a physical meaning, but for the moment, let us note that this
non-analyticity is a direct consequence of the analytic continuation
used in the imaginary time formalism. In fact, had we treated $k_{0}$
as analytically continued and calculated $\Pi(k_{0})$ in
(\ref{analytic}), we would
have obtained the non-vanishing result 
\begin{equation}
\Pi (k_{0}) = e^{2}\,{{\rm sech}^{2} {\beta m\over 2}\over 1 +
\tanh^{2} {\beta m\over 2} \tan^{2} {\beta k_{0}\over 2}}\,{\tan
{\beta k_{0}\over 2}\over k_{0}}\label{naive}
\end{equation}
Notice that this is an analytic function, apart from simple poles in
the complex $k_{0}$-plane at $k_{0} = {(2n+1)\pi\over \beta} \pm
im$. Furthermore, 
for real values of the energy, this is a well behaved function of $k_{0}$,
which vanishes for $k_{0}\rightarrow \infty$ and which,
in the limit $k_{0}\rightarrow 0$, coincides with $\Pi(k_{0} =0)$ in
(\ref{static}). This
would lead to a vanishing $\Delta$. Namely, if the external energy is not
properly analytically continued, the amplitude would not only violate the Ward
identity, but also would not lead to any non-analyticity, which has a
physical meaning. 
So, in this model, the main criterion for the correct
analytic continuation is provided by the Ward identity as well as the
causality condition (to be discussed in section {\bf IV}).
Thus, we see that all these things are quite
intricately connected and all these features carry over to higher
dimensions as we will see in the subsequent sections (in fact, these
features are present in any theory).

\section{Long wave limit Kronecker energy terms in QED}

To see that the Kronecker delta non-analyticity is not simply a
feature of the $0+1$ dimensional theory and arises in higher
dimensions as well, let us analyze $3+1$ dimensional QED in the long
wave limit. First of all, we note that, at finite temperature, the
one loop photon self-energy takes the form \cite{kalashnikov,kapusta,lebellac}
(in the imaginary time formalism)
\begin{equation}
\Pi_{\mu\nu} (k^{0},\vec{k}) = - {4e^{2}T\over (2\pi)^{3}} \sum_{n} \int
d^{3}p\;{\delta_{\mu\nu}(m^{2}+ p^{2} + p\cdot k) - p_{\mu}k_{\nu} -
p_{\nu}k_{\mu} - 2 p_{\mu}p_{\nu}\over
(p^{2}+m^{2})((p+k)^{2}+m^{2})}\label{selfenergy}
\end{equation}
where $p^{0} = (2n+1)\pi T$ and $k^{0}$ is a multiple of $2\pi T$. We
are interested in the thermal contribution which can be obtained
through the simple contour representation
\begin{equation}
T \sum_{n} f(p_{0} = i(2n+1)\pi T) =\; (T=0\;{\rm part})\; - {1\over 2\pi
i} \int_{-i\infty +\epsilon}^{i\infty + \epsilon}
dp_{0}\,(f(p_{0}+f(-p_{0}))\,{1\over e^{\beta p_{0}}+1}\label{sum1}
\end{equation}
Since the zero temperature part of the amplitude is analytic at the
origin, let us concentrate only on the thermal part of the amplitude.
 
At finite temperature, the self energy can be described in terms of
two functions $\Pi_{T}$ and $\Pi_{L}$, which are related to
$\Pi_{00},\Pi_{\mu\mu}$ as
\begin{equation}
\Pi_{L} = \left(- 1 + {k_{0}^{2}\over |\vec{k}|^{2}}\right)
\Pi_{00},\qquad \Pi_{T} = -{1\over 2}\left(\Pi_{\mu\mu} +
\Pi_{L}\right)\label{decomposition}
\end{equation}
Therefore, let us look at
$\Pi_{00}$ and $\Pi_{\mu\mu}$, which are easier to evaluate. Let us
start with the evaluation of the thermal part of $\Pi_{00}$ which has
the form (see (\ref{selfenergy}), (\ref{sum1}))
\begin{equation}
\Pi_{00}^{\rm thermal}(k_{0},\vec{k}) = {4e^{2}\over (2\pi)^{4}i} \int d^{3}p
\int_{-i\infty+\epsilon}^{i\infty+\epsilon} dp_{0}\,n_{F}(p_{0})\left[
{m^{2}+ p^{2} + p\cdot k - 2p_{0}k_{0} - 2p_{0}^{2}\over
(p^{2}+m^{2})((p+k)^{2}+m^{2})} +  (k_{0}\leftrightarrow -k_{0})\right]\label{00}
\end{equation}
where $n_{F}(p_{0})= {1\over e^{\beta p_{0}}+1}$ represents the
Fermi-Dirac 
distribution function. It is clear from this that, as in the $0+1$
dimensional model, the 
structure of the integrand for $\Pi_{00}^{\rm thermal}(k_{0},0)$ is
different depending on whether $k_{0}=0$ or $k_{0}\neq 0$. When
$k_{0}\neq 0$ the integrand has two distinct poles, while for
$k_{0}=0$, the two poles coincide, much like in the $0+1$ dimensional
example. 

For $k_{0}\neq 0$ such that $|k_{0}|<< m$, a direct evaluation leads to
\begin{equation}
\Pi_{00}^{\rm thermal} (k_{0},0) = {2e^{2}\over (2\pi)^{3}} \int
d^{3}p\,{1\over k_{0}}\left(n_{F}(E_{p}+k_{0}) -
n_{F}(E_{p})\right) + (k_{0}\leftrightarrow -k_{0})\label{analytic1}
\end{equation}
where $E_{p}=(\vec{p}^{2}+m^{2})^{1/2}$, while for $k_{0}=0$, we obtain
\begin{equation}
\Pi_{00}^{\rm thermal} (0,0) = {4e^{2}\over (2\pi)^{3}} \int
d^{3}p\,{dn_{F}(E_{p})\over dE_{p}}\label{static1}
\end{equation}
It is clear that the behavior of these quantities is completely
analogous to what we have seen in the $0+1$ dimensional model. In
particular, let us note that if we evaluate (\ref{analytic1}) by 
analytically continuing $k_{0}$, then, we would obtain
$\lim_{k_{0}\rightarrow 0}\Pi_{00}^{\rm thermal}(k_{0},0) =
\Pi_{00}^{\rm thermal}(0,0)$. Namely, in such a case, the amplitude
will be analytic, which will be in violation with the Ward identity,
as we will show. On the other hand, if we use the fact that $k_{0} =
2\ell\pi T$, then, $n_{F}(p_{0}+k_{0})= n_{F}(p_{0})$ and the result in
(\ref{analytic1}) vanishes, as is the case in the $0+1$ dimensional
model, and this is completely consistent with the Ward identity. In
this case, however, there will be a non-analyticity which 
can be related to a physical quantity as we will see.

If we define, as in the $0+1$ dimensional model (we can add the zero
temperature part as well, but it is analytic and, therefore, would drop out of
the expression),
\begin{equation}
\Delta_{00} = \Pi_{00}^{\rm thermal}(0,0) - \lim_{k_{0}\rightarrow 0}
\Pi_{00}^{\rm thermal}(k_{0},0) = \Pi_{00}^{\rm thermal} (0,0) =
{4e^{2}\over (2\pi)^{3}} \int
d^{3}p\,{dn_{F}(E_{p})\over dE_{p}}\label{disc} 
\end{equation}
then, with the proper analytic continuation that respects Ward
identities, we see that this amplitude is non-analytic and the
non-analyticity can be determined, at high temperatures, to be
\begin{equation}
\Delta_{00} \approx - {e^{2}T^{2}\over 3}\label{electric}
\end{equation}

The thermal part of the trace of the
self-energy can, similarly, be obtained from (\ref{selfenergy}) and
(\ref{sum1}) and has the form
\begin{equation}
\Pi_{\mu\mu}^{\rm thermal} (k_{0},\vec{k}) = {4e^{2}\over (2\pi)^{4}i}
\int d^{3}p \int_{-i\infty+\epsilon}^{i\infty+\epsilon}
dp_{0}\,n_{F}(p_{0}) \left[{4m^{2}+2p^{2}+2p\cdot k\over
(p^{2}+m^{2})((p+k)^{2}+m^{2})}+ (k_{0}\leftrightarrow - k_{0})\right]
\label{mumu}
\end{equation}
This integrand, too, has the structure alluded to earlier and we note,
without going into details again, that with proper analytic
continuation that respects Ward identities, we obtain, for $k_{0}\neq
0$ with $|k_{0}|<< m$,
\begin{eqnarray}
\Pi_{\mu\mu}^{\rm thermal}(k_{0},0) & = & -{2e^{2}\over (2\pi)^{3}}
\int {d^{3}p\over E_{p}}\left[{m^{2}\over
E_{p}}{1\over k_{0}}\left(n_{F}(E_{p}+k_{0}) - n_{F}(E_{p})\right) -
n_{F}(E_{p})\left({m^{2}\over E_{p}^{2}} + 2\right) +
(k_{0}\leftrightarrow -k_{0})\right]\nonumber\\
 & = &  {4e^{2}\over (2\pi)^{3}} \int {d^{3}p\over
E_{p}}\,n_{F}(E_{p}) \left({m^{2}\over E_{p}^{2}} + 2\right)
\end{eqnarray}
while for $k_{0}=0$, we obtain
\begin{equation}
\Pi_{\mu\mu}^{\rm thermal}(0,0) = - {4e^{2}\over (2\pi)^{3}} \int
{d^{3}p\over E_{p}}\left[{m^{2}\over E_{p}} {dn_{F}(E_{p})\over
dE_{p}} - n_{F}(E_{p})\left({m^{2}\over E_{p}^{2}} + 2\right)\right]
\end{equation}
This brings out the non-analytic structure clearly (which is primarily
due to the proper analytic continuation that respects Ward identity). We
can again define
\begin{equation}
\Delta_{\mu\mu} = \Pi_{\mu\mu}^{\rm thermal} (0,0) -
\lim_{k_{0}\rightarrow 0} \Pi_{\mu\mu}^{\rm thermal} (k_{0},0) =
- {4e^{2}m^{2}\over (2\pi)^{3}} \int {d^{3}p\over
E_{p}^{2}}\,{dn_{F}(E_{p})\over dE_{p}}
\end{equation}
which, at high temperature, has a  sub-leading behavior compared with
(\ref{electric}). 

Let us next turn to the Ward identities \cite{braaten1,frenkel1} and
the  physical meaning of
these non-analyticities. First of all, we note from the Ward identity
that
\begin{equation}
k_{0}\Pi_{0\mu}^{\rm thermal} (k_{0},0) = 0
\end{equation}
which would imply that
\begin{equation}
\Pi_{0\mu}^{\rm thermal} (k_{0}\neq 0,0) = 0
\end{equation}
As we have seen, this is one of the guiding relations in the proper analytic
continuation of the external energy. Second, we have already seen that
(see (\ref{disc},\ref{electric})), at high temperatures,
\begin{equation}
\Delta_{00} = \Pi_{00}^{\rm thermal} (0,0) \approx - { e^{2}T^{2}\over
3}
\end{equation}
On the other hand, let us recall that the square of the screening
length, in a plasma, is related to $\lim_{|\vec{k}|\rightarrow 0}
\Pi_{00}^{\rm thermal} (0,\vec{k})$. We will show in the next section
that, when $k_{0}=0$, amplitudes are analytic in $|\vec{k}|$ so that
\begin{equation}
 \lim_{|\vec{k}|\rightarrow 0} \Pi_{00}^{\rm thermal} (0,\vec{k}) =
 \Pi_{00}^{\rm thermal} (0,0) =  \Delta_{00} = - m_{\rm electric}^{2}
\approx - {e^{2}T^{2}\over 3}\label{electric1}
\end{equation}
This brings out the physical meaning of this non-analyticity.

Similarly, let us note from (\ref{decomposition}) that
\begin{equation}
\Pi_{T}^{\rm thermal} = - {1\over 2} \left(\Pi_{\mu\mu}^{\rm thermal} +
\left(-1 + {k_{0}^{2}\over |\vec{k}|^{2}}\right)\Pi_{00}^{\rm
thermal}\right)
\end{equation}
so that we can, correspondingly, introduce
\begin{equation}
\Delta_{T} = \Pi_{T}^{\rm thermal} (0, 0) -
\Pi_{T}^{\rm thermal} (k_{0}\rightarrow 0, 0)\label{T}
\end{equation}
Therefore, we can write
\begin{equation}
\Delta_{T} =  {1\over
2}\left(\Delta_{00} - \Delta_{\mu\mu}\right) -
\lim_{k_{0}\rightarrow 0,\vec{k}=0}\,{1\over 2}\,{k_{0}^{2}\over
|\vec{k}|^{2}}\,\Pi_{00}
\end{equation}
The last term, in the above expression, appears singular, but because of
the Ward identity, $\Pi_{00}\sim |\vec{k}|^{2}$ so that it is, in fact, well
behaved. At high temperature, this can be evaluated to give
\begin{equation}
\Delta_{T} \approx - {e^{2}T^{2}\over 9} = - {1\over 3} m_{\rm
electric}^{2} = - m_{\rm plasmon}^{2}\label{plasmon}
\end{equation}
This shows that the non-analyticity in $\Pi_{T}^{\rm thermal}$ can be
related to the plasmon mass. Let us note that the first term, on
the right hand side in (\ref{T}), is called the magnetic mass, which
vanishes to one loop order. In such a case, we recognize that
\begin{equation}
\Delta_{T} = - \Pi_{T}^{\rm thermal} (k_{0}\rightarrow 0, 0)
\end{equation}
and it can be checked, from the standard results, that it agrees with
our result in (\ref{plasmon}).

\section{General structure of Kronecker energy terms in QED}

Let us next look at the self-energy in QED away from the long wave
limit. It would seem that, for small $\vec{k}$, we can make a Taylor
expansion in the external momentum. However, when expanded in this
way, every term in the series (except the first term) becomes
divergent, when $k_{0}\rightarrow 0$, so that such an expansion does
not make  sense. On the other
hand, looking at the expressions in (\ref{00}, \ref{mumu}), we note
that, at the poles (the pole can always be chosen to be at
$p^{2}+m^{2} = 0$ or $p_{0}=  E_{p}$ with appropriate shift), the
denominator that needs to be expanded is
\begin{equation}
\left.{1\over (p+k)^{2} + m^{2}}\right|_{p_{0}= E_{p}} = {1\over
\vec{k}^{2}+2\vec{k}\cdot \vec{p} - k_{0}^{2} - 2E_{p}k_{0}}
\end{equation}
and can be naturally expanded in powers of ${|\vec{k}|\over
k_{0}}$. Therefore, let us define
\begin{equation}
s^{2} = {|\vec{k}|^{2}\over k_{0}^{2}}\label{angle}
\end{equation}
The parameter $s$, then, would represent the direction along which one
approaches the origin in the $(k_{0},|\vec{k}|)$ plane, as
$k_{0}\rightarrow 0$, with the ratio ${|\vec{k}|\over k_{0}}$ held
fixed.  For example,
$s=0$ will correspond to the long wave
limit, while $s\rightarrow \infty$ will denote the static limit. The
amplitudes can now be expressed as depending on $(k_{0},s)$ and,
correspondingly, as before, we can define
\begin{equation}
\Delta_{00}(s) = \Pi_{00}^{\rm thermal}(0,0) -
\Pi_{00}^{\rm thermal} (k_{0}\rightarrow 0,s),\qquad
\Delta_{\mu\mu} (s) = \Pi_{\mu\mu}^{\rm thermal} (0,0) -
\Pi_{\mu\mu}^{\rm thermal} (k_{0}\rightarrow 0,s)
\end{equation}

As we have seen in the last section, $\Delta_{\mu\mu}$ leads to
sub-leading contributions at high temperature and, therefore, for
simplicity,  we will confine our
discussions to $\Delta_{00}$ only. This can, in fact, be exactly
evaluated, in the high temperature limit, and has the explicit form
\begin{equation}
\Delta_{00} (s) = - {m_{\rm electric}^{2}\over 2s} \log
{1+s\over 1-s}\label{00s}
\end{equation}
This can be expanded for small values of $s$ as
\begin{equation}
\Delta_{00} (s) = - m_{\rm electric}^{2} \left(1 + {s^{2}\over 3} +
{s^{4}\over 5} + \cdots \right)
\end{equation}
In particular, it shows that when $s=0$ (namely, in the long wave limit),
this discontinuity is related to the square of the
electric  mass, as
was obtained earlier in (\ref{electric1}). Furthermore, we note that
\begin{equation}
\lim_{s\rightarrow \infty}\,\Delta_{00} (s) = 0
\end{equation}
This proves, as stated in the last section, that amplitudes are analytic
at the origin, in this plane, in the static limit.
We have already seen that $|\Delta_{00}(s=0)|$ has the physical meaning
of being the square of the screening mass. It is, therefore, tempting
to  speculate
that we can think of $|\Delta_{00} (s)|$, for small
$s$,  as the square of the 
screening mass for quasi-static electric fields at finite 
temperature. 

In the earlier sections, we have shown how the analytic continuation
of energy crucially depends on the Ward identity which, in turn, leads
to a non-analyticity. We have given a physical meaning to the
non-analyticity in a gauge theory. It is, of course, known that the
non-analyticity, at finite temperature, arises because of new branch
cuts appearing in a thermal field theory \cite{weldon,das1}. It is
because of  this, say
for example, that $\Delta_{00} (s)$ depends on the direction along
which one approaches the origin. In a sense, therefore, we
can think of the non-analyticity as a consequence of the causality in
thermal field theory. In this section, we will show how the causality
relations, in a thermal gauge theory, are further constrained by the Ward
identity to give a physical non-analyticty.

To understand these things a little better, let us simply analyze the
behavior of $\Pi_{00}^{\rm thermal}$. We note that the
amplitudes that we are evaluating, in the imaginary time formalism,
correspond to retarded amplitudes. These satisfy a dispersion relation
of the form
\begin{equation}
{\rm Re}\,\Pi_{00}^{\rm dispersion} (k_{0},s) = {\rm Re}\,
\Pi_{00}^{\rm dispersion}(k_{0},k) =  {1\over
\pi} \int_{-\infty}^{\infty} d\omega\,{{\rm
Im}\,\Pi_{00}(\omega,k)\over \omega - k_{0}} = {2\over \pi}
\int_{0}^{\infty} d\omega {\omega {\rm Im}\,\Pi_{00}(\omega,k)\over
\omega^{2} - k_{0}^{2}}\label{causality}
\end{equation}
where we have defined $k = |\vec{k}|$ and used the fact that ${\rm
Im}\,\Pi_{00}(\omega,k)$ is odd in $\omega$. At finite temperature, there
are two  branch cuts in the self-energy. The usual branch cut, which
also  occurs at zero
temperature,  satisfies $\omega^{2}-k^{2}\geq 4m^{2}$ (we
will refer to this as the time-like cut), while the new cut that arises
in thermal field theory, due to the existence of additional channels of
reaction, satisfies $\omega^{2} - k^{2}\leq 0$ (we will refer to this
as the space-like cut).

The contribution of the time-like cut is insensitive to how we
approach the origin in the $(k_{0},k)$ plane and can be obtained from the
dispersion relation to be
\begin{equation}
{\rm Re}\,\Pi_{00}^{{\rm time}-{\rm like}} (0,0) = {e^{2}\over \pi^{2}}
\int_{2m}^{\infty} d\omega\,\omega \sqrt{1-{4m^{2}\over
\omega^{2}}}\,n_{F}({\omega\over 2})
\end{equation}
At high temperature, this has the behavior
\begin{equation}
{\rm Re}\,\Pi_{00}^{{\rm time}-{\rm like}} (0,0) \approx {e^{2}T^{2}\over 3}
\end{equation}
The contribution from the space-like cut, on the other hand, depends
on how we approach the origin and has the form
\begin{equation}
{\rm Re}\,\Pi_{00}^{{\rm space}-{\rm like}} (k_{0},k) = {2\over \pi}
\int_{0}^{k} d\omega\,{\omega {\rm Im}\,\Pi_{00}(\omega,k)\over
\omega^{2} - k_{0}^{2}} = {2\over \pi} \int_{0}^{1} dv\,{v {\rm
Im}\,\Pi_{00}(vk,k)\over v^{2} - {k_{0}^{2}\over k^{2}}}
\end{equation}
where we have defined $v={\omega\over k}$ and
\begin{equation}
{\rm Im}\,\Pi_{00}(vk, k) = -{2e^{2}v\over \pi}
\int_{m\over \sqrt{1- v^{2}}}^{\infty}
d\bar{\omega}\,\bar{\omega} n_{F}(\bar{\omega})
\end{equation}
It is clear that this depends on how we approach the origin. At high
temperature, this can be evaluated to give
\begin{equation}
{\rm Re}\,\Pi_{00}^{{\rm space}-{\rm like}} (k_{0},k) \approx -
{e^{2}T^{2}\over 3} \left[1 - {1\over 2s} \log \left|{1+s\over
1-s}\right|\right]
\end{equation}
where, as before, we have identified $s={k\over k_{0}}$.

The sum of the two contributions gives, at high temperature,
\begin{equation}
{\rm Re}\, \Pi_{00}^{\rm dispersion} (k_{0},k<<m) = {\rm
Re}\,\Pi_{00}^{{\rm time}-{\rm like}} + {\rm
Re}\,\Pi_{00}^{{\rm space}-{\rm like}}  \approx {e^{2}T^{2}\over
3}\,{1\over 2s}\,\log \left|{1+s\over 1-s}\right|\label{dispersion}
\end{equation}
Let us note, however, that the expression in (\ref{dispersion}) does not
vanish when $k=0$ (namely, $s=0$), as the Ward
identity would require of the amplitude $\Pi_{00}$. In other words, the
naive dispersion relation does not automatically lead to the actual
amplitude. On the other hand,
at high  temperature,
\begin{equation}
{\rm Re}\,\Pi_{00} (k_{0},k<<m) = {\rm Re}\,\Pi_{00}^{\rm
dispersion} (k_{0},k<<m) - {e^{2}T^{2}\over 3}\label{actual}
\end{equation}
vanishes when $k=0$, in accordance with the Ward
identity. This implies that the dispersion relation, that is
consistent with the Ward identity, can be written as
\begin{equation}
{\rm Re}\,\Pi_{00} (k_{0},k) + {e^{2}T^{2}\over 3} = {1\over \pi}
\int_{-\infty}^{\infty} d\omega\, {{\rm Im}\,\Pi_{00}(\omega,k)\over
\omega - k_{0}}\label{subtracted}
\end{equation}

The result above suggests the following subtracted
dispersion relation at finite temperature,
\begin{equation}
{\rm Re}\,\left[\Pi_{00} (k_{0},k) - \Pi_{00} (0,k)\right] = {1\over
\pi} \int_{-\infty}^{\infty} d\omega\,\left({1\over \omega - k_{0}} -
{1\over \omega}\right)\,{\rm Im}\,\Pi_{00}(\omega, k)\label{subtract}
\end{equation}
There are several things to note from here. First, as
$k_{0},k\rightarrow 0$, the contribution of the time-like cut cancels
from this expression so that the entire contribution comes only from
the space-like cut. Second, ${\rm
Im}\,\Pi_{00} (0,k) =
0$, since the imaginary part is an odd function of
energy. Furthermore, if we  define
\begin{equation}
\Delta_{00} (k_{0},k) = \Pi_{00} (0,k) - \Pi_{00} (k_{0},k)
\end{equation}
then, it follows from our earlier observations that we can write
\begin{equation}
{\rm Re}\,\Delta_{00}(k_{0},k<<m) = {1\over \pi} \int_{-k}^{k}
d\omega\,\left({1\over \omega - k_{0}} - {1\over \omega}\right) {\rm
Im}\,\Delta_{00} (\omega, k)\label{subtract1}
\end{equation}
Namely, $\Delta_{00}(k_{0},k<<m)$ satisfies a simple subtracted form
of dispersion relation, where only the space-like cut
contributes. This shows that the contributions to the Kronecker delta
energy terms, which only exist at finite temperature, come from the
absorption of virtual, space-like quanta by the real particles present
in the plasma.

Relations (\ref{subtract}) and (\ref{subtract1}) can, in fact, be explicitly
checked in the simple $0+1$ dimensional model, where we can see that
they are valid at all temperatures. In this model, there is no
branch cut and the only non-analyticity results from the contribution
at $k_{0}=0$. From the Ward identity (\ref{ward}), we see that when
$k_{0}\neq 0$, both the real and the imaginary parts of the
self-energy vanish. At $k_{0}=0$, the self-energy is real (see
(\ref{static})) so that the
imaginary part vanishes, but it can be shown to vanish as
\begin{equation}
{\rm Im}\, \Pi (k_{0}) = \pi \Delta\,  k_{0} \delta
(k_{0})\label{imaginary}
\end{equation}
which is manifestly an odd function. Even though this vanishes, it can
contribute to the subtracted dispersion relation as follows.

Let us first look at the subtracted relation (\ref{subtract}) in the
$0+1$ dimensional QED
model. For $k_{0}=0$, we see that this relation is identically
satisfied. When $k_{0}\neq 0$, we have $\Pi(k_{0})=0$ so that
\begin{equation}
{1\over \pi}\int_{-\infty}^{\infty} d\omega\, \left({1\over \omega -
k_{0}} - {1\over \omega}\right)\pi \Delta\, \omega \delta (\omega) = -
\Delta = - {\rm Re}\,\Pi (0)
\end{equation}
Therefore, relation (\ref{subtract}) is exactly
satisfied. Since Im $\Pi (0) = 0$, an analogous relation to
(\ref{subtract1}), can also be seen to hold in a similar manner.

\section{Conclusion}

In this short note, we have discussed some aspects of the
non-analytic terms that arise in  thermal gauge theories. We
have shown explicitly, in the $0+1$ dimensional model, that Ward
identities  require a particular
analytic continuation of the external energy, which is responsible for
the Kronecker delta non-analyticity in the amplitudes. This analysis,
therefore, provides a further reason for such an analytic
continuation (which is conventionally used to ensure good high energy
behavior) in other theories as
well. The Kronecker delta terms are not particular to $0+1$
dimensional gauge theories. They arise in any theory and in higher
dimensions as well. In $3+1$ QED, such terms, in the long wave limit,
have a direct physical meaning, namely, they are related to the
electric mass and the plasmon mass. We have shown that the dispersion
relations, in QED, need a subtraction at finite temperature
to be compatible with the Ward identities. The Kronecker delta energy terms,
in particular, satisfy a simple subtracted dispersion relation, which
gets contributions only from the space-like cuts that arise at
finite temperature.

\vskip .5cm
\noindent{\bf Acknowledgment:}

This work was supported in part by US DOE grant number
 DE-FG-02-91ER40685 and by CNPq and FAPESP (Brasil).

\end{document}